\documentclass[twoside,twocolumn,9pt]{article}
\usepackage{extsizes}
\usepackage[super,sort&compress,comma]{natbib} 
\usepackage[version=3]{mhchem}
\usepackage[left=1.5cm, right=1.5cm, top=1.785cm, bottom=2.0cm]{geometry}
\usepackage{balance}
\usepackage{mathptmx}
\usepackage{sectsty}
\usepackage{graphicx} 
\usepackage{lastpage}
\usepackage[format=plain,justification=justified,singlelinecheck=false,font={stretch=1.125,small,sf},labelfont=bf,labelsep=space]{caption}
\usepackage{float}
\usepackage{fancyhdr}
\usepackage{fnpos}
\usepackage[english]{babel}
\addto{\captionsenglish}{%
  
}
\usepackage{array}
\usepackage{droidsans}
\usepackage{charter}
\usepackage[T1]{fontenc}
\usepackage[usenames,dvipsnames]{xcolor}
\usepackage{setspace}
\usepackage[compact]{titlesec}
\usepackage{hyperref}

\usepackage{epstopdf}

\definecolor{cream}{RGB}{222,217,201}

\begin{document}

\pagestyle{fancy}
\thispagestyle{plain}
\fancypagestyle{plain}{
\renewcommand{\headrulewidth}{0pt}
}

\makeFNbottom
\makeatletter
\renewcommand\LARGE{\@setfontsize\LARGE{15pt}{17}}
\renewcommand\Large{\@setfontsize\Large{12pt}{14}}
\renewcommand\large{\@setfontsize\large{10pt}{12}}
\renewcommand\footnotesize{\@setfontsize\footnotesize{7pt}{10}}
\makeatother

\renewcommand{\thefootnote}{\fnsymbol{footnote}}
\renewcommand\footnoterule{\vspace*{1pt}%
\color{cream}\hrule width 3.5in height 0.4pt \color{black}\vspace*{5pt}} 
\setcounter{secnumdepth}{5}

\makeatletter 
\renewcommand\@biblabel[1]{#1}            
\renewcommand\@makefntext[1]%
{\noindent\makebox[0pt][r]{\@thefnmark\,}#1}
\makeatother 
\renewcommand{\figurename}{\small{Fig.}~}
\sectionfont{\sffamily\Large}
\subsectionfont{\normalsize}
\subsubsectionfont{\bf}
\setstretch{1.125} 
\setlength{\skip\footins}{0.8cm}
\setlength{\footnotesep}{0.25cm}
\setlength{\jot}{10pt}
\titlespacing*{\section}{0pt}{4pt}{4pt}
\titlespacing*{\subsection}{0pt}{15pt}{1pt}

\fancyfoot{}
\fancyfoot[LO,RE]{\vspace{-7.1pt}\includegraphics[height=9pt]{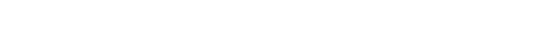}}
\fancyfoot[CO]{\vspace{-7.1pt}\hspace{13.2cm}\includegraphics{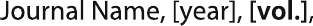}}
\fancyfoot[CE]{\vspace{-7.2pt}\hspace{-14.2cm}\includegraphics{head_foot/RF}}
\fancyfoot[RO]{\footnotesize{\sffamily{1--\pageref{LastPage} ~\textbar  \hspace{2pt}\thepage}}}
\fancyfoot[LE]{\footnotesize{\sffamily{\thepage~\textbar\hspace{3.45cm} 1--\pageref{LastPage}}}}
\fancyhead{}
\renewcommand{\headrulewidth}{0pt} 
\renewcommand{\footrulewidth}{0pt}
\setlength{\arrayrulewidth}{1pt}
\setlength{\columnsep}{6.5mm}
\setlength\bibsep{1pt}

\makeatletter 
\newlength{\figrulesep} 
\setlength{\figrulesep}{0.5\textfloatsep} 

\newcommand{\topfigrule}{\vspace*{-1pt}%
\noindent{\color{cream}\rule[-\figrulesep]{\columnwidth}{1.5pt}} }

\newcommand{\botfigrule}{\vspace*{-2pt}%
\noindent{\color{cream}\rule[\figrulesep]{\columnwidth}{1.5pt}} }

\newcommand{\dblfigrule}{\vspace*{-1pt}%
\noindent{\color{cream}\rule[-\figrulesep]{\textwidth}{1.5pt}} }

\makeatother

\twocolumn[
  \begin{@twocolumnfalse}
{\includegraphics[height=30pt]{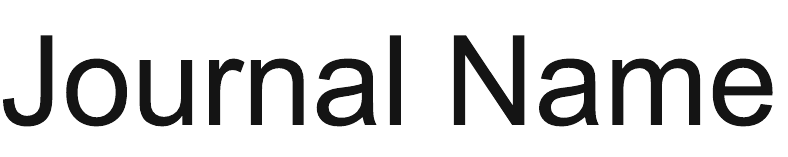}\hfill\raisebox{0pt}[0pt][0pt]{\includegraphics[height=55pt]{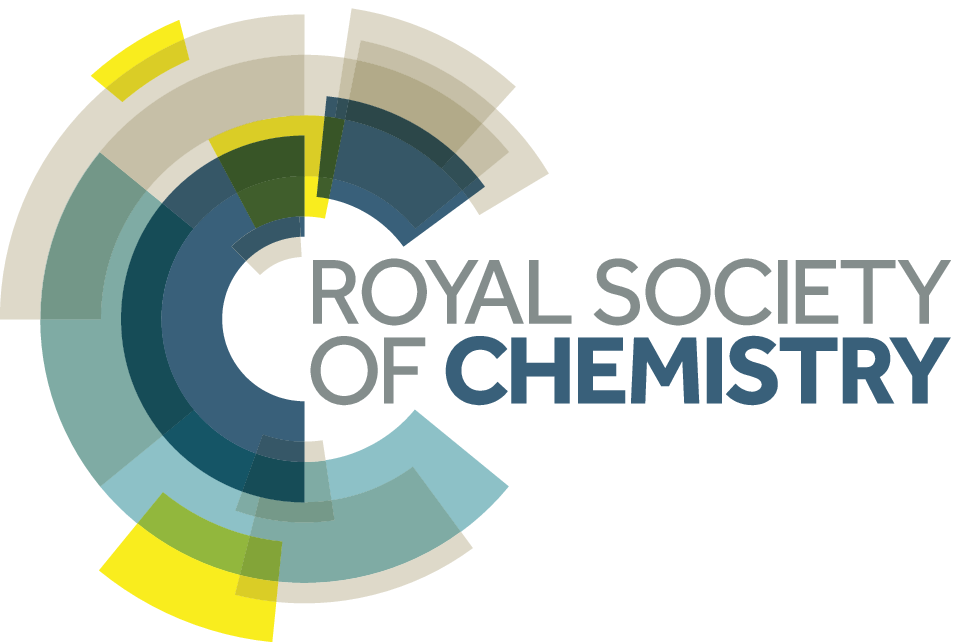}}\\[1ex]
\includegraphics[width=18.5cm]{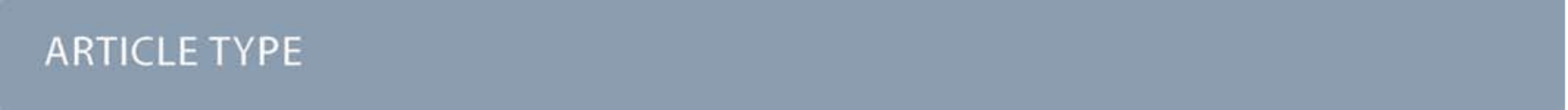}}\par
\vspace{1em}
\sffamily
\begin{tabular}{m{4.5cm} p{13.5cm} }

\includegraphics{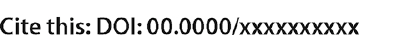} & \noindent\LARGE{\textbf{On the Thermomechanical Properties and Fracture Patterns of the Novel Nonbenzenoid Carbon Allotrope (Biphenylene Network): A Reactive Molecular Dynamics Study$^\dag$}} \\
\vspace{0.3cm} & \vspace{0.3cm} \\

 & \noindent\large{M. L. Pereira J\'unior,\textit{$^{a}$} W. F. da Cunha,\textit{$^{a}$} R. T. de Sousa Junior,\textit{$^{b}$} G. D. Amvame Nze,\textit{$^{b}$} D. S. Galv\~ao,\textit{$^{c,d}$} and L. A. Ribeiro J\'unior $^{a,\ast}$} \\

\includegraphics{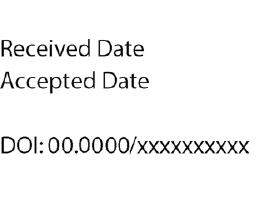} & \noindent\normalsize{Recently, a new two-dimensional carbon allotrope, named biphenylene network (BPN) was experimentally realized. The BPN structure is composed of four-, six-, and eight-membered rings of sp$^2$-hybridized carbon atoms. In this work, we carried out fully-atomistic reactive (ReaxFF) molecular dynamics simulations to study the thermomechanical properties and fracture patterns of non-defective and defective (nanocracks) BPN. Our results show that under uniaxial tensile loading, BPN is converted into four distinct morphologies before fracture starts. This conversion process is dependent on the stretching direction. Some of the formed structures are mainly formed by eight-membered rings, which have different shapes in each morphology. In one of them, a graphitization process was observed before the complete fracture. Importantly, in the presence of nanocracks, no new morphologies are formed. BPN exhibits a distinct fracture process when contrasted to graphene. After the critical strain threshold, the graphene transitions from an elastic to a brittle regime, while BPN can exhibit different inelastic stages. These stages are associated with the appearance of new morphologies. However, BPN shares some of the exceptional graphene properties. Its calculated Young's modulus and melting point values are comparable to the graphene ones, about 1019.4 GPa and 4024K, respectively.} \\

\end{tabular}

 \end{@twocolumnfalse} \vspace{0.6cm}

  ]

\renewcommand*\rmdefault{bch}\normalfont\upshape
\rmfamily
\section*{}
\vspace{-1cm}


\footnotetext{\textit{$^{a}$~Institute of Physics, University of Bras\'ilia, 70910-900, Bras\'ilia, Brazil.}}
\footnotetext{\textit{$^{b}$~Department of Electrical Engineering, University of Bras\'{i}lia 70919-970, Brazil.}}
\footnotetext{\textit{$^{c}$~Applied Physics Department, University of Campinas, Campinas, S\~ao Paulo, Brazil.}}
\footnotetext{\textit{$^{d}$~Center for Computing in Engineering and Sciences, University of Campinas, Campinas, S\~ao Paulo, Brazil.}}
\footnotetext{\textit{$^{\ast}$~Corresponding Author: ribeirojr@unb.br}}




\section{Introduction}

Since the discovery of graphene \cite{geim2010rise}, carbon-based 2D systems have attracted considerable attention from the scientific community due to unique properties and potential applications, such as novel energy storage \cite{ma2017carbon,miao2020recent,geng2020structure} and conversion \cite{zheng2015engineering,latibari2018carbon,zhang2015carbon}, with good efficiency and low environmental impact \cite{hu2018carbon,zhao2019carbon}. Moreover, they often exhibit a very useful combination of low cost and relatively easy fabrication (controllable synthesis, which can result in different structures \cite{kumar2018recent,novoselov2012roadmap,li2008graphene,fan2021biphenylene,toh_N}). These features make carbon-based 2D systems suitable for such a wide range of applications \cite{liu2016carbon}. 

The advent of graphene renewed the interest in 2D carbon and recently other structures have been proposed in the literature but most of them have not been yet experimentally realized \cite{enyashin2011graphene,zhang2015penta,wang2015phagraphene,li2017psi,wang2018popgraphene,zhuo2020me,chen2020pai}. Some of these novel 2D allotropes have the advantage of presenting a non-zero bandgap, which is one of the limiting features of using graphene in a series of optoelectronic applications. 

Very recently, a new synthetic route to obtain a graphene-like structure --- 2D biphenylene network (BPN) \cite{fan2021biphenylene} --- was reported. BPN consists of the periodic arrangement of a set of four, six, and eight carbon rings fused side by side. Armchair-like BPN nanoribbons present similar electronic properties to armchair graphene nanoribbons \cite{fan2021biphenylene}. In this sense, the 2D layered BPN structure emerges as a promising material that needs to be further investigated in terms of structural, electronic, and  thermomechanical properties.

\begin{figure*}[!t]
	\centering
	\includegraphics[width=\linewidth]{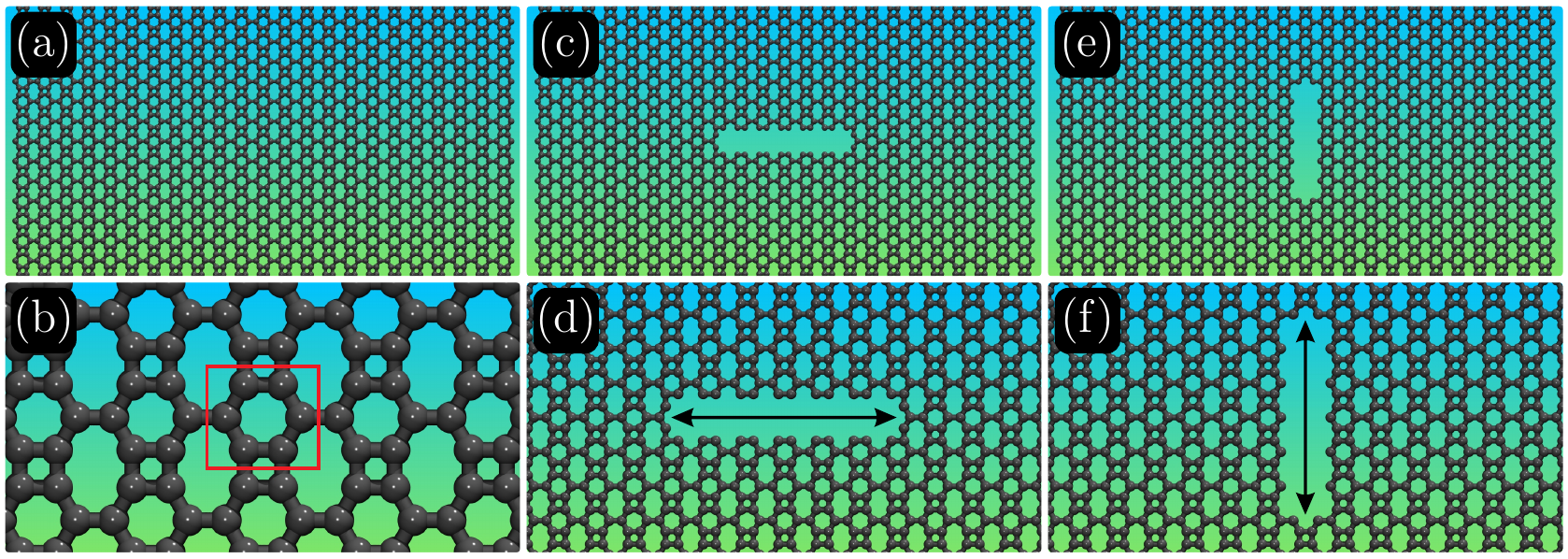}
	\caption{Schematic representation of the BPN structures studied here: (a,b) non-defective BPN, (c,d) a BPN with a horizontally aligned nanocrack ($x$-direction), and (e,f) a BPN with a vertically aligned nanocrack ($y$-direction). The bottom sequence of panels shows zoomed regions of the corresponding lattices. The red rectangle in panel (b) represents the asymmetric unity cell.}
	\label{fig:systems}
\end{figure*}
When it comes to the synthesis of nanostructures, the presence of defects is a reality \cite{gleiter1992nanostructured}. In many cases, such a trend should not be considered a failure in the process but an important feature \cite{fang2015defect}. For instance, defects are known to provide important changes in the mobility of charge carriers in otherwise semiconducting systems, such as armchair graphene nanoribbons \cite{DACUNHA2015171}. Defects are also a reality in organic synthesis. However, to take advantage of possible structural defects, a detailed study of their influence on the system performance is needed. 

Due to BPN promising features, there are several theoretical works investigating their structural and electronic properties, most of them carried out before their experimental realization \cite{hosseini2020theoretical,esfandiarpour2019density,rahaman2016structural,baughman1987structure}. These works have addressed different BPN features, such as their electronic structure  \cite{rahaman2017metamorphosis,bafekry2021biphenylene} and their potential use as a material source for Lithium-ion batteries \cite{ferguson2017biphenylene}. Nevertheless, detailed studies of their thermomechanical properties, and their fracture patterns, have not been yet investigated. More specifically, an understanding of the influence of lattice defects (such as cracks/notches) is still missing in the literature. 

In this work, we have carried out fully atomistic reactive (Reaxff) molecular dynamics simulations to investigate the thermomechanical properties of both pristine and defective (with nanocracks/notches) BPN. We were able to map different phase transitions regarding the response of the system structure when subjected to uniaxial strain. We also show that BPN exhibits a quite distinct mechanical behavior in comparison to graphene. After a critical strain threshold, the graphene sheet goes directly from elastic to brittle regimes, while BPN has different intermediate structural stages. These stages are associated with the formation of other kinds of structures and/or linear atomic changes. Interestingly, we found that the calculated BPN Young's modulus value and melting point are comparable to the graphene ones. 

\section{Methodology}   

The mechanical and thermal properties of non-defective and defective BPN were investigated using fully atomistic MD simulations with the reactive force field ReaxFF \cite{ashraf2017extension}, as implemented in the large-scale atomic/molecular massively parallel simulator (LAMMPS) code \cite{plimpton_JCP}. Importantly, the ReaxFF potential allows the formation and breaking of chemical bonds during the dynamics, which is necessary to investigate the fracture mechanisms \cite{senftle2016reaxff}. The BPN studied here are illustrated in Figure \ref{fig:systems}. Figures \ref{fig:systems}(a,b), \ref{fig:systems}(c,d), and \ref{fig:systems}(e,f)  present the non-defective BPN, a BPN with a horizontally aligned nanocrack, and a BPN with a vertically aligned nanocrack, respectively. 

In Figure \ref{fig:systems}, the bottom sequence of panels shows zoomed regions of the corresponding lattices. We used BPN structural models with dimensions of $96.56\times 95.58$ \AA${^2}$ with the atoms at the edges fixed to avoid spurious effects during the stretching processes. The horizontally and vertically aligned nanocracks have dimensions of $23.30\times 5.70$ \AA${^2}$ and $27.00\times 5.30$ \AA${^2}$, respectively. The total number of carbon atoms is 3450 for the non-defective case, 3414 for the horizontal nanocrack, and 3416 for the vertical nanocrack, respectively.  

We numerically integrated the equations of motion using the velocity-Verlet integrator with a time-step of $0.05$ fs. The tensile stress was applied considering a uniaxial strain along with the non-periodic $x$ and $y$ directions, for an engineering strain rate of $1.0 \times 10^{-6}$ fs$^{-1}$. Before the BPN stretching, the structures were equilibrated within an NPT ensemble at a constant temperature of 300K and null pressures using a Nos\'e-Hoover thermostat \cite{hoover1985canonical} during 200 ps. More details about this methodological approach can be found in the reference \cite{junior2020thermomechanical}.

\begin{figure*}[!h]
	\centering
	\includegraphics[width=\linewidth]{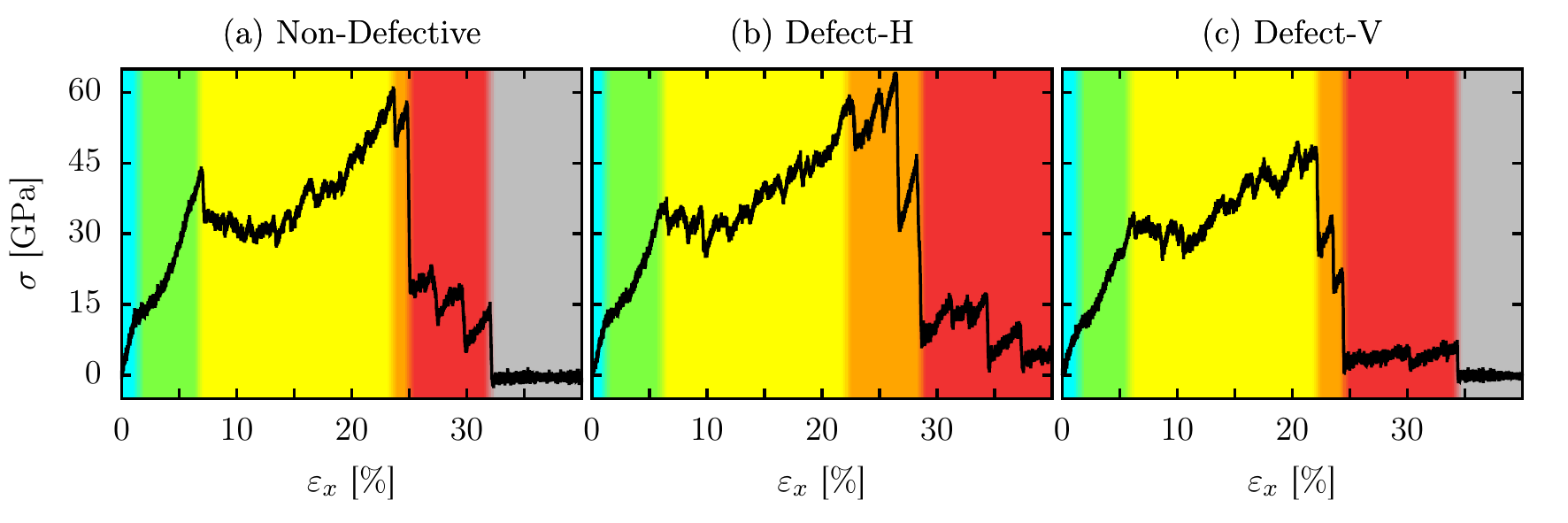}
	\caption{Stress-strain curves (for strain applied along the x-direction) for: (a) non-defective, (b) horizontally defective (Defect-H), and (c) vertically defective (Defect-V) BPN.}
	\label{fig:ss_curves_x}
\end{figure*}

\begin{figure*}[!t]
	\centering
	\includegraphics[width=\linewidth]{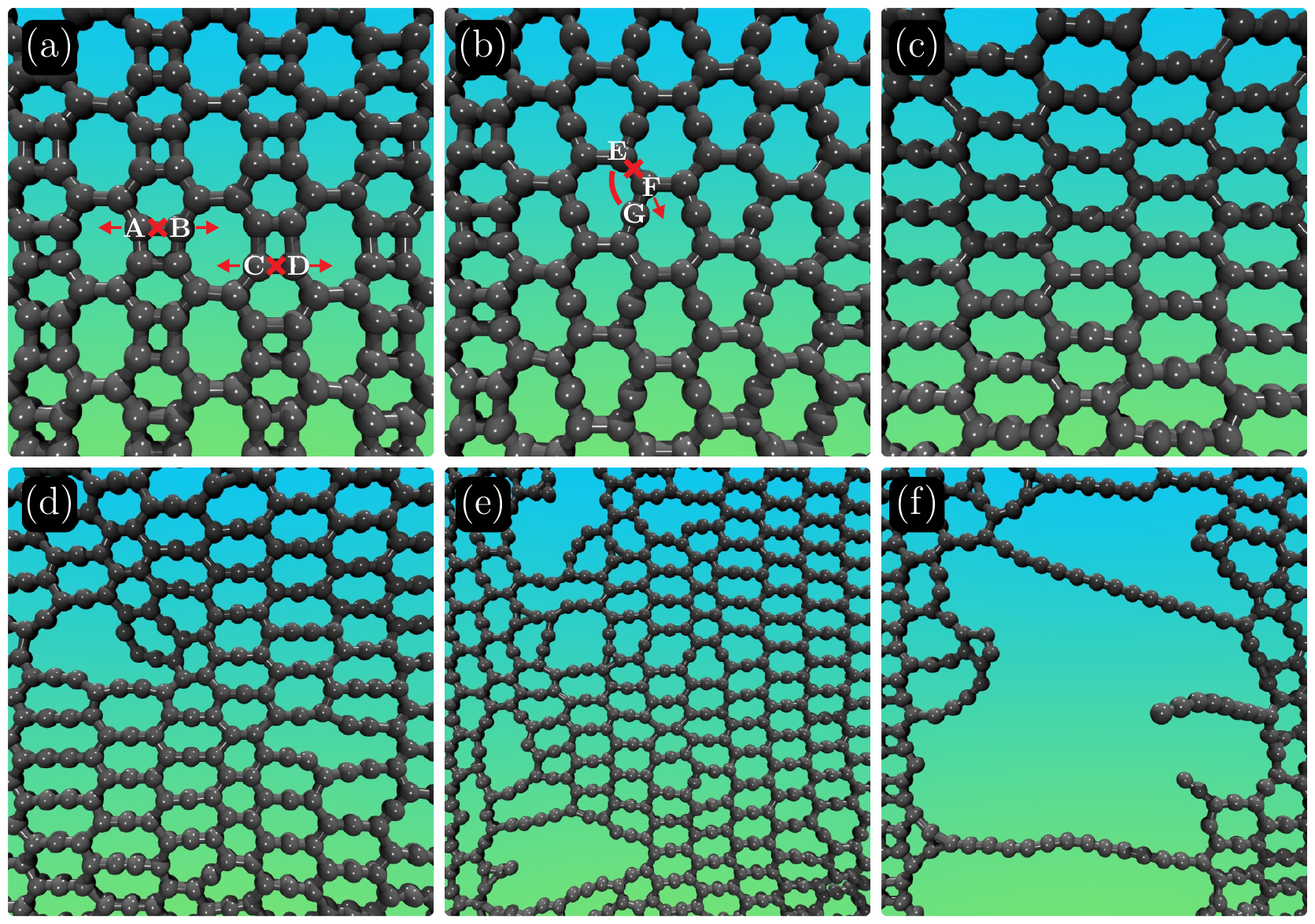}
	\caption{Representative MD snapshots of the BPN at different strain values (applied along the x-direction, $\varepsilon_x$) for the non-defective case. The panels in this figure represent the following strain percentages: (a) 0\%, (b) 5\%, (c) 20\%, (d) 23\%, (e) 28\%, and (f) 30\%. See text for discussions.}
	\label{fig:frat-mech-x}
\end{figure*}

The BPN structures shown in Figure \ref{fig:systems} were continuously stretched up to their complete structural failure (fracture), which can be identified by the fractured and irreversible patterns presented by the lattices and by the abrupt change in the stress values. The maximum applied strain was 50\%. We calculated the von Mises stress (VM) per-atom values \cite{mises_1913}. These values provide information on the fracture process once they help to locate the fracture point or region. More details about the VM calculations can be found in the LAMMPS manuals \cite{lammps_manual} and in reference \cite{junior2020thermomechanical}. 
Using this MD simulation scheme, we obtained the following elastic properties from the stress-strain curves: Young’s modulus ($Y_M$), Fracture Strain (FS), and Ultimate Strength (US). The MD snapshots and trajectories were obtained using the visualization and analysis software VMD \cite{HUMPHREY199633}.

\section{Results}

We begin our discussion by investigating the stress ($\sigma$) response as a function of the x-direction applied strain ($\varepsilon_x$), as illustrated in Figure \ref{fig:ss_curves_x}. The x-strain corresponds to the horizontal direction in Figure \ref{fig:systems}, i.e., parallel to the nanocrack of Figure \ref{fig:systems} (d). The following observations are valid for the three represented systems. As we can see from this figure, as the strain increases, the unstressed regime (cyan regions) of the BPN is followed by the green and yellow regions, respectively. These regions are related to the elastic regime and the first and second inelastic regimes, respectively. A further stress increase yields a new lattice configuration that corresponds to the orange region. This region is associated with the beginning of the structural failure (fracture) process. Ultimately, linear atomic chains (LACs) are formed until the complete BPN fracture (red regions). Completely fractured lattices are in the gray region, in which the stress value drops to zero. 

\begin{figure*}[!h]
	\centering
	\includegraphics[width=0.92\linewidth]{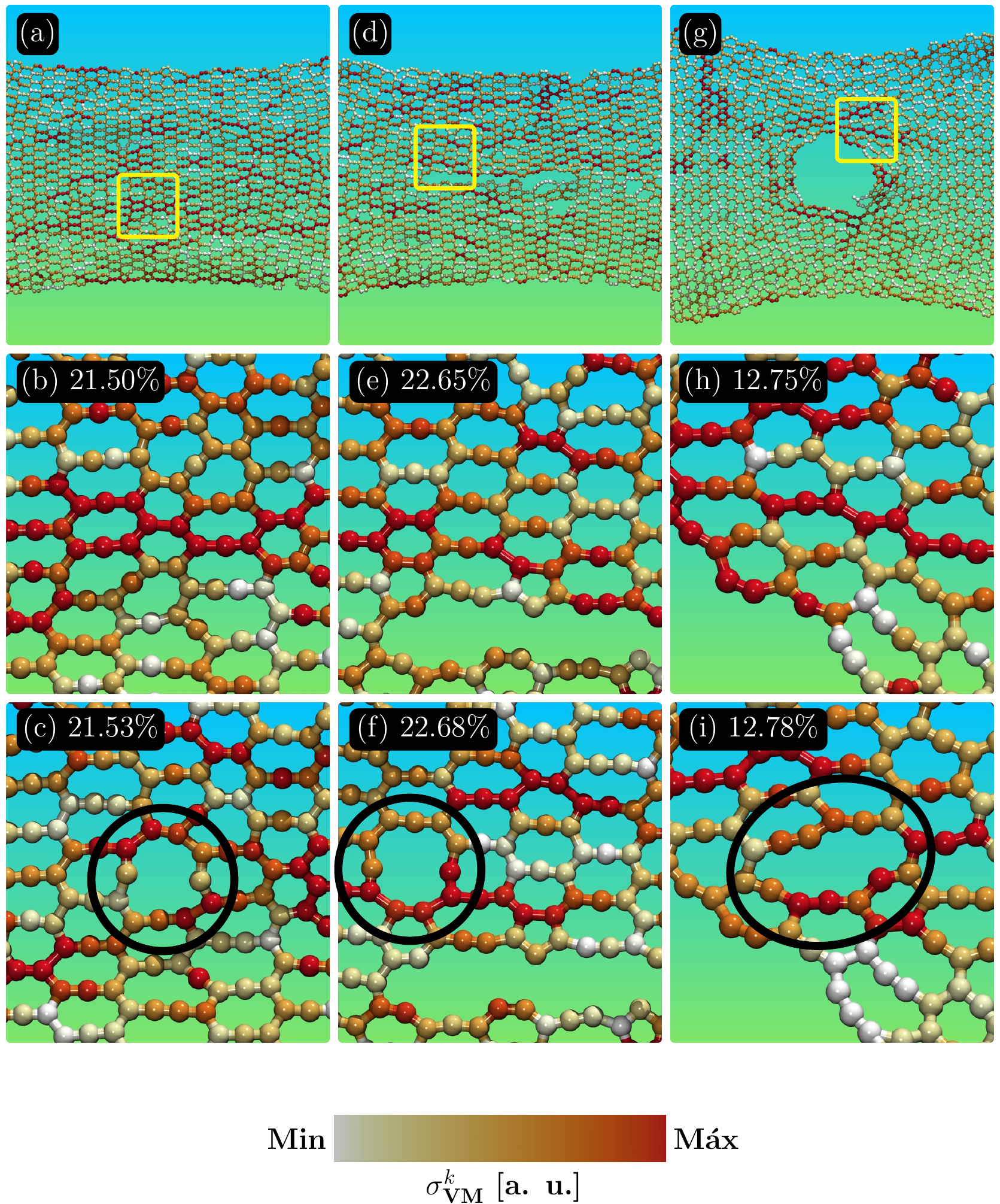}
	\caption{Representative MD snapshots with the von Mises stress ($\sigma^{k}_\text{VM}$) per-atom values, for the BPN  under a tensile loading applied along the x-direction. Figures \ref{fig:VM-x}(a-c), \ref{fig:VM-x}(d-e), and \ref{fig:VM-x}(g-i) present the $\sigma^{k}_\text{VM}$ distribution for the non-defective BPN (at 21.50 \% and 21.53 \% of strain before and after the fracture, respectively), Defect-H BPN (at 22.65 \% and 22.68 \% of strain before and after fracture, respectively), and Defect-V BPN (at 12.75 \% and 12.78 \% of strain before and after the fracture, respectively), respectively. The top sequence of panels shows a general view of the BPN cases for the time step before their first fracture regime. The middle sequence of panels is the zoomed regions indicated by the yellow squares. These regions show the first bond breaking for each case. The bottom sequence of panels illustrates the bond reconstructions and the newly formed rings (black circles) after the first bond breaking cycle.}
	\label{fig:VM-x}
\end{figure*}

The three panels of Figure \ref{fig:ss_curves_x} represent the behavior of BPN with different structures: \ref{fig:ss_curves_x}(a) the non-defective structure, \ref{fig:ss_curves_x}(b) the structure with a horizontal nanocrack defect (Defect-H), and \ref{fig:ss_curves_x}(c) the structure with a vertical nanocrack defect (Defect-V). Although the structural fracture process starts approximately at the same strain values, the process as a whole is distinct. The nanocrack leads to the appearance of more resilient LACs, as can be seen from the broadening of the red regions of Figures \ref{fig:ss_curves_x}(b) and \ref{fig:ss_curves_x}(c), when compared to Figure \ref{fig:ss_curves_x}(a). The complete fracture now occurs at higher strain values. Also, the Defect-H case is the most resilient.       

Figures \ref{fig:frat-mech-x}(a-f) show representative MD snapshots for the strain applied along the horizontal direction ($\varepsilon_x$) for the non-defective case. In the Supplementary Material, we present the video for this MD simulation (see video1.mpg). Figure \ref{fig:frat-mech-x}(a) shows that, in the initial stages of the process, the A and B type atoms begin to have their mutual bonds dissociated, thus causing the first phase transition in the lattice structure. As the strain increases, the E-G atoms come closer together, which leads to a bond formed between them. As a result, this further weakens the bonds between the E and F atoms, thus starting the new phase of the inelastic regime. We observed that each new-formed phase is less stressed than the previous one: the structure as a whole is structurally softened. This trend occurs because as $\varepsilon_x$ increases, the porosity of the structure increases as vacancy defects resulting from the new bond pattern appears, as illustrated in Figure \ref{fig:frat-mech-x}(d). The starting of the structural failure (fracture) is shown in Figure \ref{fig:frat-mech-x}(e), it further continues in Figure \ref{fig:frat-mech-x}(f) with the LAC formation until the BPN is completely fractured. 

To better discuss the BPN dynamics under a tensile loading applied along the x-direction, we present in Figure \ref{fig:VM-x} representative MD snapshots with the von Mises stress ($\sigma^{k}_\text{VM}$) per-atom values \cite{mises_1913,junior2020thermomechanical}. The ($\sigma^{k}_\text{VM}$) values provide local structural information on the fracture process since they can indicate the regions from which the fracture starts. In this way, Figures \ref{fig:VM-x}(a-c), \ref{fig:VM-x}(d-e), and \ref{fig:VM-x}(g-i) present the $\sigma^{k}_\text{VM}$ distribution for the non-defective BPN (at 21.50 \% and 21.53 \% of strain before and after the fracture, respectively), Defect-H BPN (at 22.65 \% and 22.68 \% of strain before and after fracture, respectively), and Defect-V BPN (at 12.75 \% and 12.78 \% of strain before and after the fracture, respectively). Such values for the fracture strain express the aforementioned fact that the vertical defect tends to further weaken the lattice when compared to the non-defective and horizontally cracked systems. The top sequence of panels shows a general view of the BPN cases for the time step before their first fracture stage. The middle sequence of panels shows the regions within the yellow squares for their corresponding cases, which were zoomed for clarity. These regions show the first bond breaking for each case. The bottom sequence of panels illustrates the bond reconstructions and the newly formed rings (black circles) after the first bond breaking. 

In Figure \ref{fig:VM-x}(a-c) and \ref{fig:VM-x}(d-f) we can see that $\sigma^{k}_\text{VM}$ tends to accumulate in the carbon-carbon bonds almost parallel to the direction of the applied strain, as can be inferred from the bonds in red in Figures \ref{fig:VM-x}(a) and \ref{fig:VM-x}(d). As a consequence, the bonds parallel to the strain direction that connect the two eight-membered rings (i.e, the bonds belonging to the rings with six atoms) are the first ones to break (see Figures \ref{fig:VM-x}(b) and \ref{fig:VM-x}(e)). After the first bond breaking, bond reconstructions lead to the formation of new rings (or pores) connecting 11 carbon atoms, as highlighted by the black circles in Figures \ref{fig:VM-x}(c) and \ref{fig:VM-x}(f). It is worthwhile to mention that these bond breaking/reconstruction mechanisms occur in several regions of the structure at the same time. In the Supplementary Material, we present the videos for this MD simulation (see video1.mpg and video2.mpg). 

In the case of Defect-H BPN, the nanocrack is aligned to the direction of the applied strain. In this sense, during the stretching process, the atoms within the defective region come closer when the strain rate is increased. Such as process allows bond reconstructions within the nanocrack, which are responsible for slightly increasing the strain rate experienced by the BPN (and a changing in BPN degree of stiffness) for this case when contrasted to the non-defective one. When it comes to the case in which the nanocrack is aligned perpendicularly to the strain direction (Defect-V BPN), we can see that most bonds that accumulate stress are not parallel to the strain direction, as shown in Figures \ref{fig:VM-x}(g-i). Interestingly, the vertical nanocrack leads to an almost uniform distribution of the accumulated stress, and the Defect-H BPN has only a small number of bonds in red (see Figure \ref{fig:VM-x}(g)). In this case, the first bonds to break are the ones nearby the nanocrack (see Figure \ref{fig:VM-x}(h)). The bonds that connect two eight-membered rings side-by-side are the first to break, yielding a pore composed of 13 carbon atoms. In all cases, the first bond-breaking results in an abrupt fracture followed by fast crack propagation. 

In the Supplementary Material, we present videos for an overall visualization of the whole fracture process of these cases (see video1.mpg, video2.mpg, and video3.mpg). In these videos, we can clearly see that the nanocracks preclude the above-mentioned BPN phase transitions. 

\begin{figure*}[!t]
	\centering
	\includegraphics[width=\linewidth]{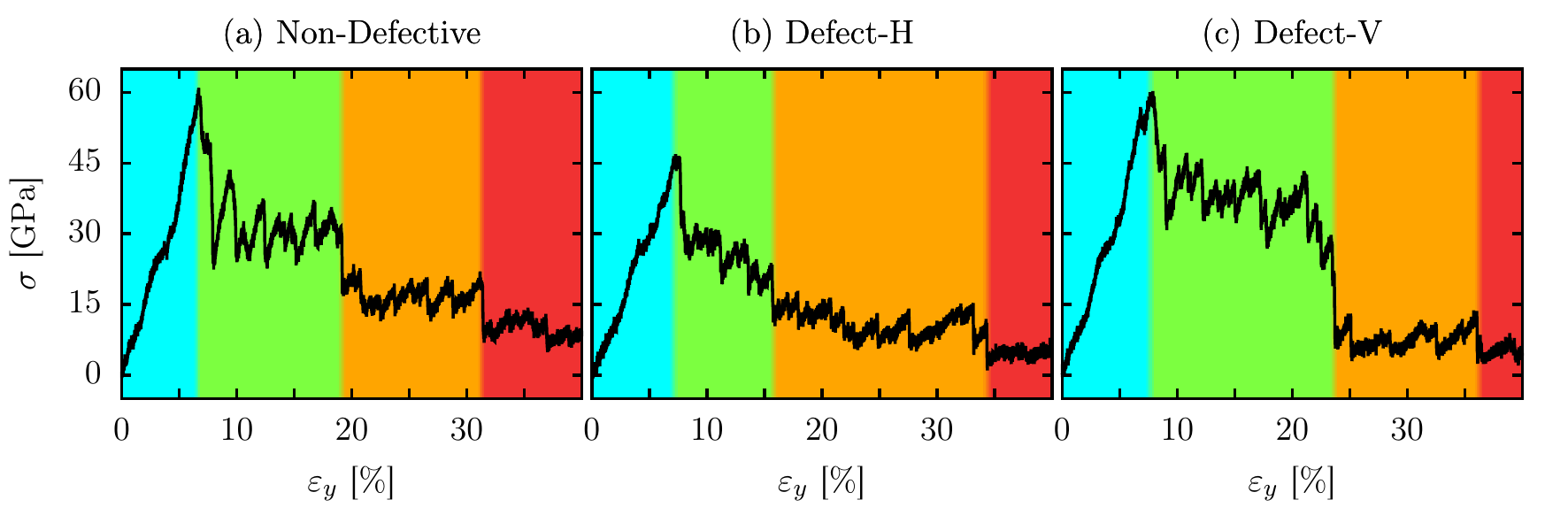}
	\caption{Stress-strain curves (for strain applied along the y-direction) relation for: (a) non-defective, (b) horizontally defective (Defect-H), and (c) vertically defective (Defect-V) BPNs.}
	\label{fig:ss_curves_y}
\end{figure*}

\begin{figure*}[!t]
	\centering
	\includegraphics[width=\linewidth]{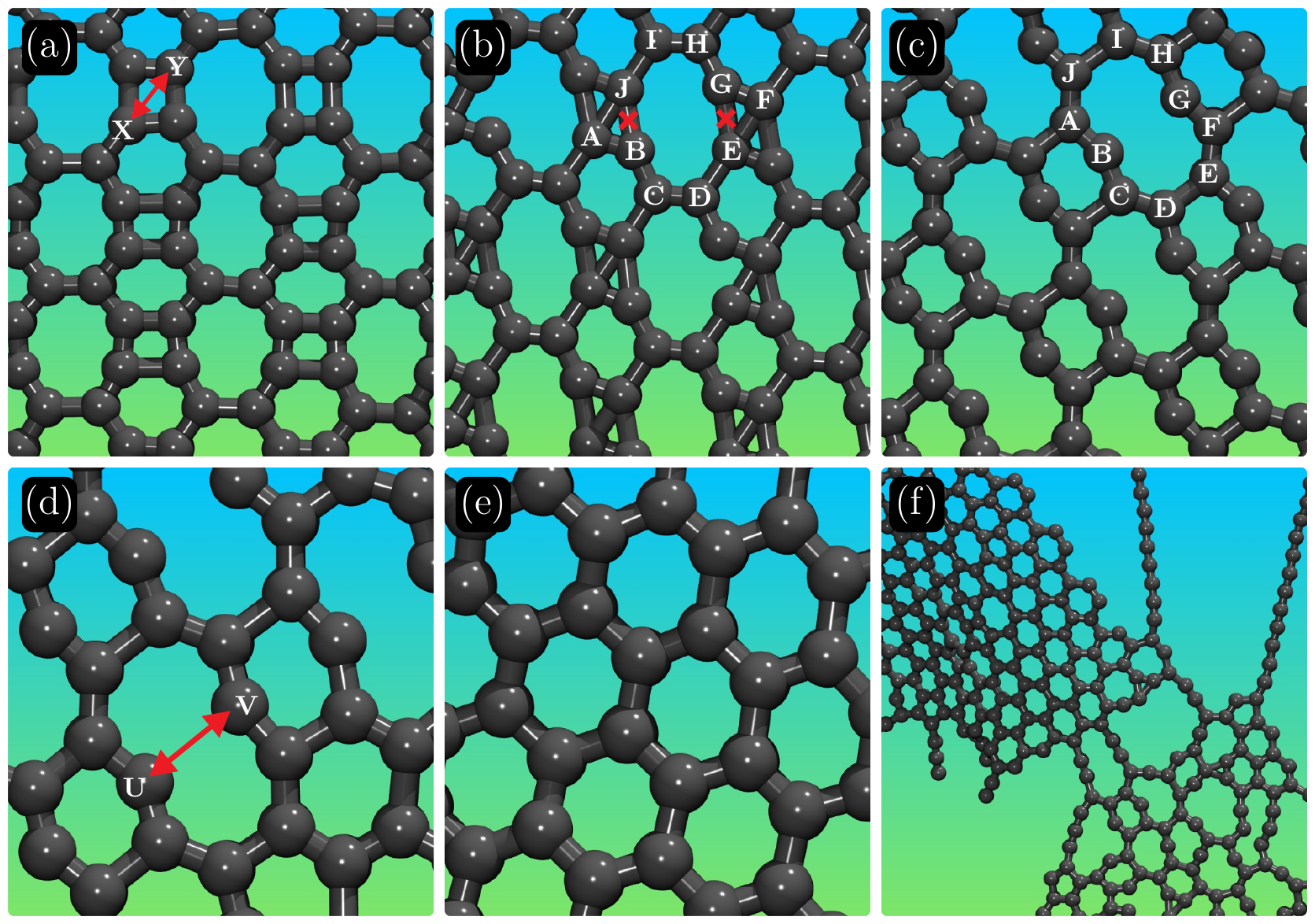}
	\caption{Representative MD snapshots of the BPN at different strain values (applied along the y-direction, ($\varepsilon_y$)) for the non-defective case. The panels in this figure represent the following strain percentages: (a) 0\%, (b) 7\%, (c) 8\%, (d) 10\%, (e) 20\%, and (f) 25\%. See text for discussions.}
	\label{fig:frat-mech-y}
\end{figure*}

\begin{figure*}[!h]
	\centering
	\includegraphics[width=0.92\linewidth]{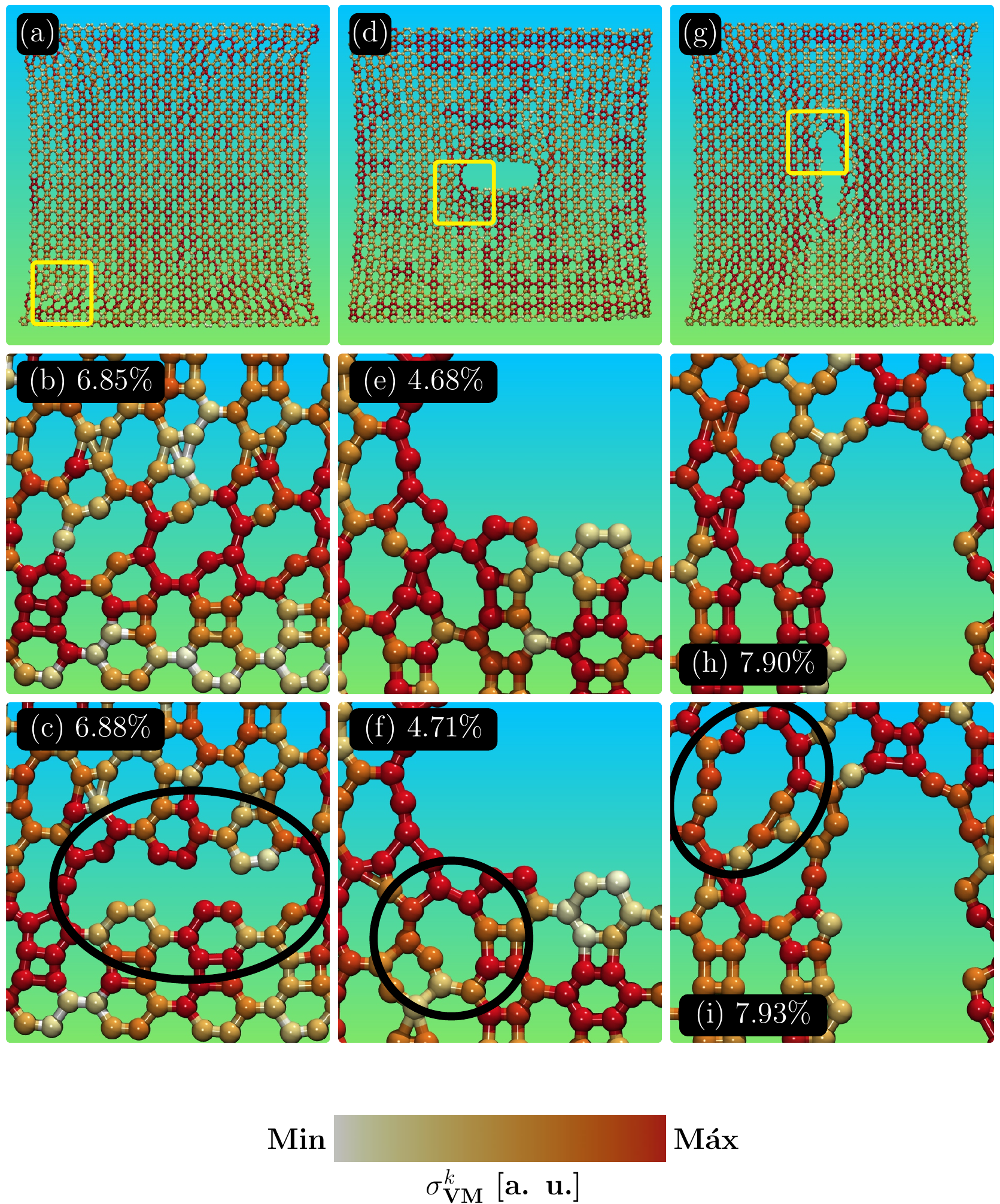}
	\caption{Representative MD snapshots with the von Mises stress ($\sigma^{k}_\text{VM}$) per-atom values, for the BPN under a tensile loading applied along the y-direction. Figures \ref{fig:VM-x}(a-c), \ref{fig:VM-x}(d-e), and \ref{fig:VM-x}(g-i) present the $\sigma^{k}_\text{VM}$ distribution for the non-defective BPN (at 6.85 \% and 6.88 \% of strain before and after the fracture, respectively), Defect-H BPN (at 4.68 \% and 4.71 \% of strain before and after the fracture, respectively), and Defect-V BPN (at 7.90 \% and 7.93 \% of strain before and after the fracture, respectively). The top sequence of panels shows a general view of the BPN cases for the time step before their first fracture regime. The middle sequence of panels is the zoomed regions indicated by the yellow squares. These regions show the first bond breaking for each case. The bottom sequence of panels illustrates the bond reconstructions and the newly formed rings (black circles) after the first bond-breaking cycle.}
	\label{fig:VM-y}
\end{figure*}

We now analyse the stress ($\sigma$) response as a function of the strain applied along the y-direction ($\varepsilon_y$), as shown in Figure \ref{fig:ss_curves_y}. In the Supplementary Material, we present the video for this MD simulation (see video4.mpg). In this case, the strain is along the vertical direction, parallel to the vacancy defect of Figure \ref{fig:systems}(f). Four phases are identifiable from Figures \ref{fig:ss_curves_y}(a), \ref{fig:ss_curves_y}(b), and \ref{fig:ss_curves_y}(c) which, again, correspond to the non-defective, horizontally defective (Defect-H), and vertically defective (Detect-V) BPN, respectively. The cyan region corresponds to the non-stressed structure. At the end of this region, a phase transition process leading to structural failure begins. Green and orange regions correspond to the continuing of such a process. In red, the formation of LACs takes place, followed by the completely fractured structure (the null stress grey regions in Figure \ref{fig:ss_curves_x}, not shown in Figure \ref{fig:ss_curves_y}). In all cases, the fracture occurs at approximately the same range between 7 and 8\% of strain. The horizontal defect in Figure \ref{fig:ss_curves_y}(b) leads to an increased softened structure, which results in a slightly smaller Ultimate Stress value when compared to the other cases (about 15 GPa of difference). We can see that the vertical defect results in an increase in the structural strength. 

In Figure \ref{fig:frat-mech-y} we present representative MD snapshots that illustrate the fracture mechanism corresponding to the process described in Figure \ref{fig:ss_curves_y}. In Figure \ref{fig:frat-mech-y}(a), we can see that atoms X and Y become closely bonded. From Figure \ref{fig:frat-mech-y}(b), we observe that the J-B and the G-E pairs begin to dissociate, thus starting a new phase shown in Fig. \ref{fig:frat-mech-y}(c). Such a bond reconfiguration trend appears in several parts of the structure easing the fracture process in some (but not all) of those parts. In the cases that such fracture does not take place, the U and V atoms of Figure \ref{fig:frat-mech-y}(d) come closer to one other, thus creating a local graphene-type structure. This new rearrangement takes place in several regions of the structure. The system tends to become increasingly more similar to graphene, as can be seen in Figure \ref{fig:frat-mech-y}(e). In Figure \ref{fig:frat-mech-y}(f), we can observe larger pores, more graphene-type regions, and the appearance of LACs that precedes the complete fracture. Because of this dynamical behavior, this fracture pattern can be described as a \textit{tearing-like process} \cite{witkowska2011modelling}.

\begin{figure}[!t]
	\centering
	\includegraphics[width=1.0\linewidth]{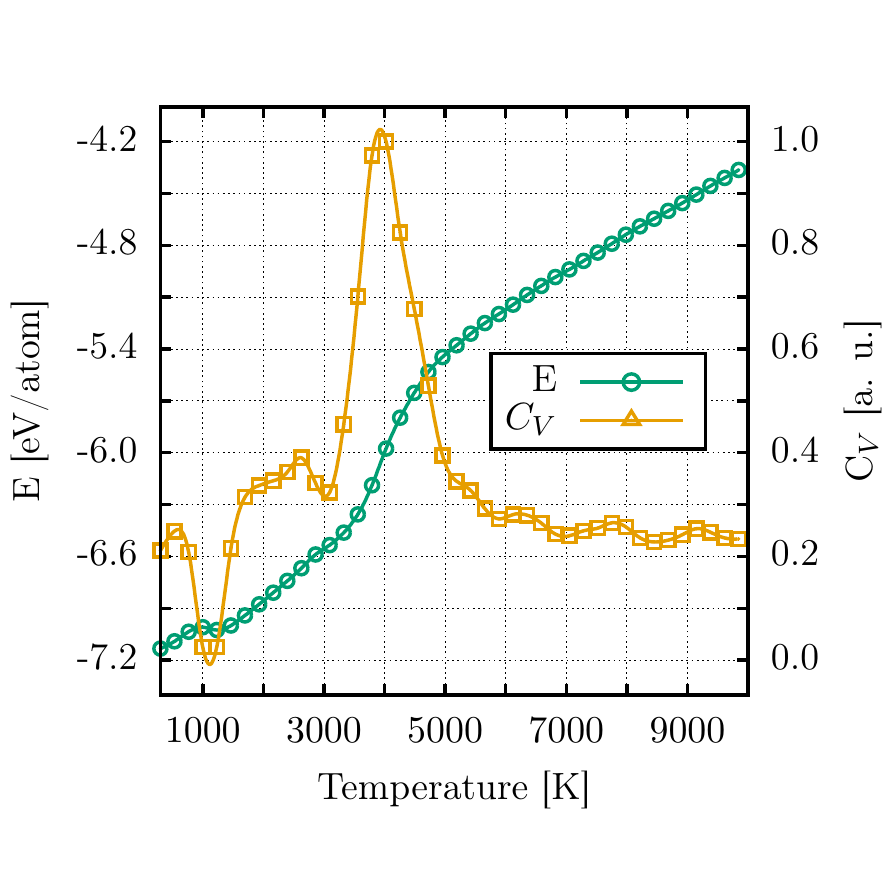}
	\caption{BPN total energy and heat capacity ($C_V$) as a function of temperature for the 250 ps heating ramp simulations (melting process).}
	\label{fig:melting_curves}
\end{figure}

\begin{figure*}[!t]
	\centering
	\includegraphics[width=\linewidth]{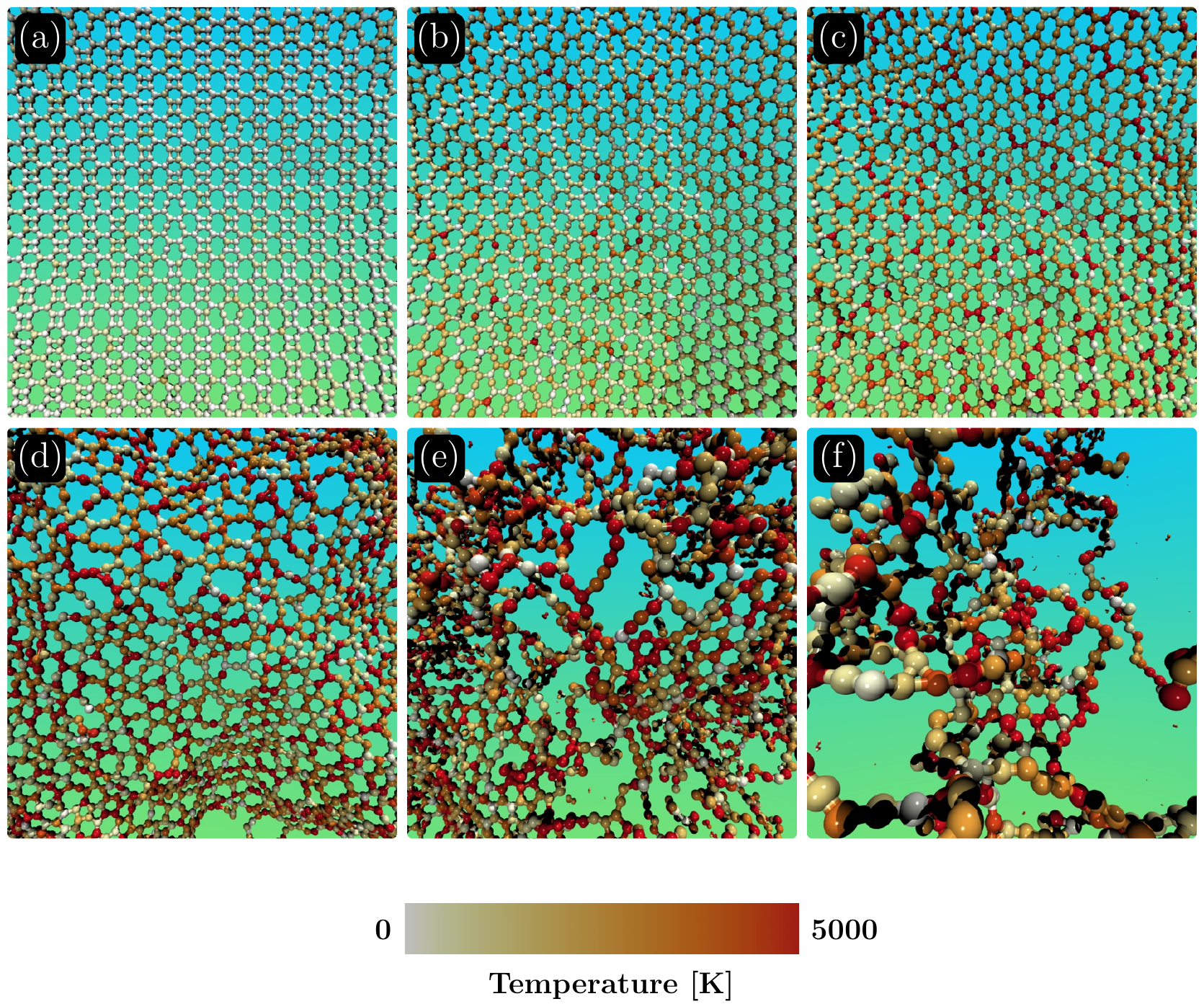}
	\caption{Representative MD snapshots for the heating ramp simulations (melting process). The color scale indicated the temperature per atom ranging from 300K up to 5000K, which are represented by the white and red colors, respectively. (a) 300K at 5 ps, (b) 1000K at 5 ps, (c) 2000K at 50 ps, (d) 3000K at 75 ps, (e) 4000K at 100 ps, and (f) 5000K at 125 ps.}
	\label{fig:melting_snaps}
\end{figure*}

In Figure \ref{fig:VM-y} we present representative MD snapshots with ($\sigma^{k}_\text{VM}$) per-atom values for the cases that consider the strain applied along the y-direction.  In Figures \ref{fig:VM-y}(a-c), \ref{fig:VM-y}(d-e), and \ref{fig:VM-y}(g-i) we present the $\sigma^{k}_\text{VM}$ distribution for the non-defective BPN (at 6.85 \% and 6.88 \% of strain before and after the fracture, respectively), Defect-H BPN (at 4.68 \% and 4.71 \% of strain before and after the fracture, respectively), and Defect-V BPN (at 7.90 \% and 7.93 \% of strain before and after the fracture, respectively). This Figure has the same layout used in Figure \ref{fig:VM-x}. In the Supplementary Material, we present videos for an overall visualization of the whole fracture process of the cases mentioned above (see video4.mpg for the non-defective case, video5.mpg for the Defect-H case, and video6.mpg for the Defect-V case, respectively). Unlike the cases of Figure \ref{fig:VM-x}, we observe a tendency for stress accumulation into the four- and six-membered rings (see Figures \ref{fig:VM-y}(a,d,g)), but not necessarily in the bonds aligned to the direction of the applied strain. This tendency disappears when the critical strain value is reached, as a result of the morphological changes (see Figures \ref{fig:VM-y}(b,e,h)). For all cases, BPN present a significant fracture strain value when contrasted to the cases of Figure \ref{fig:VM-x}. This trend is due to the fact that, in this stretching mode, the bonds break forming large pores, resembling nanocracks (see Figure \ref{fig:VM-y}(c)) or they broke near the nanocrack (see Figures \ref{fig:VM-y}(f) and \ref{fig:VM-y}(i)), which is the most fragile structural region.     

The numeric values of the elastic properties are presented in Table \ref{table:elasprop}. The Young Modulus ($Y_M$) values were calculated using a linear fitting of the stress within the first 1\% of strain. The Fracture Strain (FS) is obtained from the strain percentage that corresponds to the largest stress value, i.e., the Ultimate Strength (US). We can see that the fracture strain for the non-defective structure (23.6 \%) lies between the ones for the Defect-H (26.4 \%) and Defect-V (20.5 \%) BPN when the strain is applied along the x-direction. For the cases in which the strain is applied along the y-direction, the non-defective BPN presented the smallest FS (about 6.7 \%). The nanocracks confer to the structure a degree of porosity that tends to make it more deformable, thus increasing the FS value needed to achieve the fracture.

\begin{table*}[!h]
	\centering
	\begin{tabular}{lccclcccl}
		\hline
		\multicolumn{1}{l}{} &                                                                                                  \multicolumn{7}{c}{\phantom{\bigg|}Non-Defective\phantom{\bigg|}}                                                                                                  & \multicolumn{1}{l}{} \\ \cline{2-8}
		\multicolumn{1}{l}{} &                                     \multicolumn{3}{c}{ $\sigma_x$}                                      & \multicolumn{1}{l}{} &                                      \multicolumn{3}{c}{$\sigma_y$}                                      & \multicolumn{1}{l}{} \\ \cline{2-4}\cline{6-8}
		\multicolumn{1}{l}{} & \multicolumn{1}{c}{$Y_M$ {[}GPa{]}} & \multicolumn{1}{c}{FS {[}\%{]}} & \multicolumn{1}{c}{US {[}GPa{]}} & \multicolumn{1}{l}{} & \multicolumn{1}{c}{$Y_M$ {[}GPa{]}} & \multicolumn{1}{c}{FS {[}\%{]}} & \multicolumn{1}{c}{US {[}GPa{]}} & \multicolumn{1}{l}{} \\ 
		\multicolumn{1}{l}{} &     \multicolumn{1}{c}{1019.4}      &    \multicolumn{1}{c}{23.6}    &    \multicolumn{1}{c}{61.3}     & \multicolumn{1}{l}{} &     \multicolumn{1}{c}{745.5}      &    \multicolumn{1}{c}{6.7}    &    \multicolumn{1}{c}{61.0}     & \multicolumn{1}{l}{} \\ 
	\end{tabular}
	\begin{tabular}{lccclcccl}
		\hline
		\multicolumn{1}{l}{} &                                                                                                  \multicolumn{7}{c}{\phantom{\bigg|}Defective-H\phantom{\bigg|}}                                                                                                  & \multicolumn{1}{l}{} \\ \cline{2-8}
		\multicolumn{1}{l}{} &                                     \multicolumn{3}{c}{ $\sigma_x$}                                      & \multicolumn{1}{l}{} &                                      \multicolumn{3}{c}{$\sigma_y$}                                      & \multicolumn{1}{l}{} \\ \cline{2-4}\cline{6-8}
		\multicolumn{1}{l}{} & \multicolumn{1}{c}{$Y_M$ {[}GPa{]}} & \multicolumn{1}{c}{FS {[}\%{]}} & \multicolumn{1}{c}{US {[}GPa{]}} & \multicolumn{1}{l}{} & \multicolumn{1}{c}{$Y_M$ {[}GPa{]}} & \multicolumn{1}{c}{FS {[}\%{]}} & \multicolumn{1}{c}{US {[}GPa{]}} & \multicolumn{1}{l}{} \\ 
		\multicolumn{1}{l}{} &     \multicolumn{1}{c}{986.2}      &    \multicolumn{1}{c}{26.4}    &    \multicolumn{1}{c}{64.4}     & \multicolumn{1}{l}{} &     \multicolumn{1}{c}{570.9}      &    \multicolumn{1}{c}{7.3}    &    \multicolumn{1}{c}{46.9}     & \multicolumn{1}{l}{} \\ 
	\end{tabular}
	\begin{tabular}{lccclcccl}
		\hline
		\multicolumn{1}{l}{} &                                                                                                   \multicolumn{7}{c}{\phantom{\bigg|}Defective-V\phantom{\bigg|}}                                                                                                   & \multicolumn{1}{l}{} \\ \cline{2-8}
		\multicolumn{1}{l}{} &                                     \multicolumn{3}{c}{ $\sigma_x$}                                      & \multicolumn{1}{l}{} &                                      \multicolumn{3}{c}{$\sigma_y$}                                      & \multicolumn{1}{l}{} \\ \cline{2-4}\cline{6-8}
		\multicolumn{1}{l}{} & \multicolumn{1}{c}{$Y_M$ {[}GPa{]}} & \multicolumn{1}{c}{FS {[}\%{]}} & \multicolumn{1}{c}{US {[}GPa{]}} & \multicolumn{1}{l}{} & \multicolumn{1}{c}{$Y_M$ {[}GPa{]}} & \multicolumn{1}{c}{FS {[}\%{]}} & \multicolumn{1}{c}{US {[}GPa{]}} & \multicolumn{1}{l}{} \\ 
		\multicolumn{1}{l}{} &      \multicolumn{1}{c}{760.7}      &    \multicolumn{1}{c}{20.5}    &    \multicolumn{1}{c}{49.8}     & \multicolumn{1}{l}{} &      \multicolumn{1}{c}{693.1}      &    \multicolumn{1}{c}{7.9}    &    \multicolumn{1}{c}{60.3}     & \multicolumn{1}{l}{} \\ \hline
	\end{tabular}
	\caption{Elastic properties --- Young's Modulus ($Y_M$), Fracture Strain (FS), and Ultimate Strength (US) --- obtained by fitting the stress-strain curves (Figures \ref{fig:ss_curves_x} e \ref{fig:ss_curves_y}) for the Biphenylene Network cases investigated here.}
	\label{table:elasprop}
\end{table*}

Importantly, when the tensile loading direction is perpendicular to the nanocrack, this defect plays the role of softening the structure. The US values for Defective-V under $\varepsilon_x$ and Defective-H under $\varepsilon_y$ are 49.8 GPa and 46.9 GPa, respectively. They are the smallest values among all the structures within the same simulation protocol. On the other hand, in the simulations in which the tensile loading direction is parallel to the nanocrack, this defect increases (or does not substantially affect) the BPN resilience to tension. The US values for Defective-H under $\varepsilon_x$ and Defective-V under $\varepsilon_y$ are 64.4 GPa and 60.3 GPa, respectively. When the nanocrack is parallel to the stretching direction, the distance between the atoms at the edge of this defect decreases when the strain increases, allowing bond reconstructions within the defective region. 

These bond reconstructions are responsible for increasing the BPN US value with the nanocrack. We obtained $Y_M$ values ranging from 570.9 to 1019.4 GPa. These values are comparable to the Young's modulus of graphene, obtained through density functional theory, molecular dynamics simulations, continuum models, and finite element methods (between 698.0 and 1367 GPa) \cite{memarian2015graphene,rahaman2017metamorphosis,mortazavi2013thermal}, consistent with the experiments values, which varies from 890 GPa \cite{zhang2012measurements} up to 1000 GPa \cite{lee2008measurement}.  

Finally, we analyzed the BPN thermal stability. First, we carried out heating ramp simulations (melting process), with temperatures varying from 300K up to 10000K during 250 ps. Figure \ref{fig:melting_curves} illustrates the total energy (green) and heat capacity (yellow) as a function of temperature for the melting process of BPN. From this figure, we can see that the total energy increases quasi-linearly with the temperature with three different regimes well-defined by the different slopes in Figure \ref{fig:melting_curves}: between 300K-1000K, 1100K-3900K, and 5000K-10000K. 

In the first heating regime (300K-1000K), BPN still maintains its structural integrity. Beyond 1000K, the thermal vibrations lead to considerable changes in the original BPN morphology, resulting in a structure similar to the one presented in Figure \ref{fig:frat-mech-x}(b). This structural phase transition is characterized by the first peak and discontinuity in the $C_V$ and total energy curves, respectively. The second heating regime (between 1100K-3900K) is associated with continuous heating up of the newly formed configuration. The BPN melting takes place between 3900K-5000K. In this temperature interval, we can see the appearance of a well-pronounced peak in the $C_V$ curve and a discontinuity in the total energy curve that is related to a phase transition from a solid to a gas-like phase. The BPN melting point occurs at 4024K, which is represented by the well-pronounced peak in the $C_V$ curve. Importantly, this value is comparable to the melting point for the monolayer graphene (4095K) and for the amorphous monolayer graphene (3626K) \cite{felix2020mechanical}. Previous works have also predicted a melting point for graphene between 4000K and 6000K \cite{los2015melting,ganz2017initial,fomin2020comparative}. 

The complete BPN melting is reached at around 5000K. Up to this critical value, the total energy curve does not show changes in the slope. The slope change observed between 3900K-5000K is related to a gain in kinetic energy due to the higher atom velocities in the gas-like phase. Moreover, harmonic and torsional energies in the solid phase are converted into kinetic energy during the melting process, contributing to the increase in the total energy value. Above 5000K, the gas-phase dominates characterizing the third heating regime between 5000K-10000K.   

In Figure \ref{fig:melting_snaps} we present representative MD snapshots for the heating ramp simulations, with temperature varying from 300K up to 5000K, as previously described. The color scheme denotes the temperature per atom ranging from 300K to 5000K, which are represented by the white and red colors, respectively. At low temperatures (up to 300K), BPN presents a morphology very similar to the one of minimum energy, as illustrated in Figure \ref{fig:melting_snaps}(a) for 300K. At 1000K (see Figures \ref{fig:melting_snaps}(b) and \ref{fig:melting_snaps}(c)), the thermal fluctuations induce changes in the morphology, as mentioned above. The resulting structures are similar to the one presented in Figure \ref{fig:frat-mech-x}(b). In this sense, temperature and strain have a similar impact on the structural changes at low-temperature regimes. At 3000K, the melting process of the lattice starts to occur, as illustrated in Figure \ref{fig:melting_snaps}(d). In this figure, one can observe that the temperature effects also induce a graphitization process. Above 4000K, the thermal vibrations favor the formation of LACs, and no BPN fragments can be observed, as shown in Figure \ref{fig:melting_snaps}(a). For temperatures about 5000K, only isolated atoms and small LACs are observed and the state at the end of the heating ramp simulation is a gas-like phase.

\section{Summary and Conclusions}

We have carried out fully atomistic reactive (ReaxFF) molecular dynamics (MD) simulations to investigate the thermomechanical stability and fracture dynamics of non-defective and defective (differently oriented nanocracks, vertically/Defect-V and horizontally/Defect-H oriented) 2D biphenylene networks (BPN), which was recently synthesized \cite{fan2021biphenylene}. Our results showed that, as a general trend, under uniaxial tensile loading BPN is converted into distinct morphologies before its complete structural failure (fracture). As the strain increases, we observed the existence of three different phases in the inelastic regimes. When the critical strain (fracture strain) is reached and the fracture process starts, linear atomic chains (LAC) are formed until BPN is completely fractured.

In one of the observed inelastic regimes, a graphitization (structural transformation to graphene-like structure) process occurs before the BPN complete fracture. In the presence of nanocracks, no new structure is formed (i.e., there is no formation of distinct elastic regimes). The fracture strains obtained here were 23.6 \% (6.7 \%), 26.4 \% (7.3 \%), and 20.5 \% (7.9 \%) for the non-defective BPN, Defect-H, and Defect-V, respectively, considering the strain applied along the x-direction (y-direction). In the x-direction stretching cases, the stress tends to be accumulated in the bonds parallel to the strain direction. The vertical defect tends to further weaken the lattice when compared to the non-defective and horizontally cracked systems. In the case of Defect-H, the nanocrack is aligned to the direction of the applied strain. During the stretching process, the atoms within the defective region come closer when the strain rate is increased. This process allows bond reconstructions within the nanocrack, which are responsible for slightly increasing the strain rate experienced by the BPN structure for this case when contrasted to the non-defective one. 

Unlike the cases where the strain was applied along the x-direction, for the y-direction tensile loading, there is a tendency for stress accumulation in the four- and six-membered carbon rings, but not necessarily in the bonds aligned to the direction of the applied strain. This tendency disappears when BPN achieves its critical strain because of the morphological changes induced by the applied stress. The BPN, when stretched along the y-direction, presents a much lower fracture strain in comparison to the cases of the strain along the x-direction. For the y-direction stretching, the bonds break forming large pores, resembling new nanocracks or they are broken near the initial nanocrack, which is the most fragile structural region.     

BPN exhibits distinct fracture dynamics when contrasted to graphene. After the critical strain threshold, graphene directly transitions from an elastic to a brittle regime (\cite{felix2020mechanical}), while BPN exhibits different deformable stages. As mentioned above, these stages are associated with the formation of other structural morphologies. 

We obtained Young's modulus values ranging from 570.9 to 1019.4 GPa. These values are comparable to graphene values, obtained from other theoretical works (between 698.0 and 1367 GPa) \cite{memarian2015graphene,rahaman2017metamorphosis,mortazavi2013thermal}, and are consistent with the experimental values, which varies from 890 GPa \cite{zhang2012measurements} up to 1000 GPa \cite{lee2008measurement}.   

From the MD thermomechanical simulations we obtained that the BPN melting point occurs at 4024K. This value is comparable to the one for pristine (4095K) and amorphous graphene (3626K) \cite{felix2020mechanical}. Previous works have also predicted a melting point for graphene between 4000K and 6000K  \cite{los2015melting,ganz2017initial,fomin2020comparative}. The BPN complete melting was observed around 5000K. 

The recently reported BPN synthesis \cite{fan2021biphenylene} opens new and exciting perspectives for 2D carbon-based materials. BPN shares some of the exceptional graphene properties. We hope the present work will stimulate further BPN investigations.

\section*{Acknowledgements}
The authors gratefully acknowledge the financial support from Brazilian research agencies CNPq, FAPESP, and FAP-DF. M.L.P.J gratefully acknowledges the financial support from CAPES grant 88882.383674/2019-01. L.A.R.J acknowledges the financial support from a Brazilian Research Council FAP-DF and CNPq grants $00193.0000248/2019-32$ and $302236/2018-0$, respectively. L.A.R.J acknowledges CENAPAD-SP for providing the computational facilities. L.A.R.J. gratefully acknowledges the financial support from IFD/UnB (Edital $01/2020$) grant $23106.090790/2020-86$. The authors acknowledge the National Laboratory for Scientific Computing (LNCC/MCTI, Brazil) for providing HPC resources of the SDumont supercomputer, which have contributed to the research results reported within this paper. URL: \url{http://sdumont.lncc.br}.DSG acknowledges the Center for Computational Engineering and Sciences at Unicamp for financial support through the FAPESP/CEPID Grant \#2013/08293-7. 



\balance


\bibliography{rsc} 
\bibliographystyle{rsc} 

\end{document}